\documentstyle[12pt]{article}
\begin{document}
\thispagestyle{empty}
\renewcommand{\thefootnote}{\dagger}

N preprint.

\begin{center}
{\large\bf On the possible experimental manifestations \\
of \\
the torsion field at low energies\\}
\vskip 2truecm

\renewcommand{\thefootnote}{\ddagger}

{\bf V.G.Bagrov\footnote[1]{ e-mail adress:
bagrov@fftgu.tomsk.su\\}}

{\sl Physical Department, Tomsk State University, Tomsk, 634050,Russia.\\}

\renewcommand{\thefootnote}{\ddagger}

{\bf I.L.Buchbinder\footnote[2]{ e-mail adress:
josephb@tspi.tomsk.su\\}}

{\sl Tomsk State Pedagogical Institute, Tomsk, 634041,Russia.\\}

\renewcommand{\thefootnote}{\ddagger}

{\bf I. L. Shapiro\footnote[3]{ e-mail adress:
shapiro@fusion.sci.hiposhima-u.ac.jp\\}}

{\sl Tomsk State pedagogical Institute, 634041, Tomsk, Russia\\}

{\sl and Department of Physics, Hiroshima University, Higashi-Hiroshima, 724,
Japan.\\}

\end{center}

\vskip 2.5truecm

\noindent
{\bf Abstract.}
We construct the theoretical base
for the search of the possible experimental manifestations of the
torsion field at low energies. The weakrelativistic approximation to
 the Dirac equation in an external torsion field is considered.
For the sake of generality we introduce the external electromagnetic
field in parallel. The generalized (due to torsion dependent terms)
Pauly equation contains new terms which have the different structure
if compared with standard electromagnetic ones.
Just the same takes place for the weakrelativistic equations for spin
$\frac{1}{2}$ particle in an external torsion and electromagnetic
field. It is given the brief description of the possible experiments.

\setcounter{page}1

\renewcommand{\thefootnote}{\arabic{footnote}}
\setcounter{footnote}0
\newpage

\section{Introduction}

The modern gravitational theories are based on the geometrical description of
the gravitational field. For instance in the framework of General Relativity
the space - time is the Riemanien Manifold and the gravitational field occures
as a metric tensor field on this manifold.

It is known for a long time that the non - Riemanien Geometry gives the
appropriate base for the new gravitational theories. In particular the theory
where the torsion field is caled for the description of the gravitational
field along with metric is of special interest. In the framework of
General Relativity only the Energy - Momentum Tensor of matter fields is the
sourse of gravity. At the same time in the theories with torsion one can
consider the Spin Tensor of matter as an additional sourse of the
gravitational interaction [1,2]. The torsion field naturally arises within the
gauge approach to gravity [3,4]. Thus this kind of theories possesses
better conceptual features, and is interesting for investigation.
One can find the review of gravity with torsion in Ref's [1,2,5 - 8].
 Some questions related with our subject, namely the equation for particle with
spin $\frac{1}{2}$ have been
discussed in the papers [15 - 17].

If the torsion really exist, the investigation of it's coupling with matter
fields is of crusial importance for the understanding of this phenomena.
The interaction of free matter fields with external torsion field have
been considered in a number of papers (see, for example,
[3 - 6, 9 - 11,18,19] and
references therein). Some aspects of the interacted matter fields theory
in an external gravitational field with torsion have been discussed in [12,13]
(see also [6]). As it was shown in [7,8], the requirement
of the multiplicative
renormalizability makes us to introduce the nonminimal interaction of torsion
with spinor and scalar fields. The  renormalization
 group analysis of GUT's in an external gravitational field with
 torsion shows that the
interaction of matter fields with torsion increase in a strong gravitational
field. Therefore one can conclude that the torsion have more essential
manifestations at high energy level. On the other hand the interaction
with torsion is weakened at low energies.
This fact gives the possible reason to the
absence of torsion in a modern experimental data. Note that the low - energy
manifestations of torsion are quite interesting. In particular, the
investigation
of the weakrelativistic limit for the spin 1/2 field in an
external torsion field gives some new predictions which may turn out to
be the base for the experimental search of torsion [14].

Does the torsion field really exist? The definite answer can be obtained only
on the experimental basis. The purpose of the present paper is to consider the
theoretical grounds for the experimental tests which can detect the possible
torsion effects.

Since the torsion field is the element of the gravitational interaction, this
field must couple with matter in a universal way. Therefore we can suppose that
torsion interact with all the particles which have the nontrivial spin.
The investigation of the torsion - matter coupling is usually argued by the
possible cosmological applications. Here we discuss the possible low - energy
manifestations of torsion field, and show that the torsion field may lead to
some phenomena in the microscopic physics. Of course we do not claim the
existance of torsion, but only consider the way to test this fact.

The paper is organized as following. In section 2 we shall write the action of
spinor Dirac field in an external gravitational field with torsion.
We introduce the nonminimal interaction of  torsion with spinor field, that is
the only way to obtain consistent quantum theory [12,13] (see also [7]).
In section 3
the weakrelativistic approximation to the Dirac
equation in an external torsion and electromagnetic
fields is constructed. The generalized
(due to torsion dependent terms) Pauly equation contains new terms which
are different from  standard electromagnetic ones.
In section 4 the quasiclassical equations of motion for the
weakrelativistic particle with spin $\frac{1}{2}$ in an external torsion
and electromagnetic fields is derived. These equations contains the standard
terms corresponding to the interaction with electromagnetic field and also some
new terms related with torsion. In spite of usual point of view we find that
this terms have the different structure if compared with electromagnetic ones.
We use these new terms in section 5 where the brief description of the
possible experiments is given.

\section{Spinor field in an external gravitational field with torsion}

Let us start with the basic notations for the gravity with torsion.
In the space - time with metric $g_{\mu\nu}$ and torsion
$T^\alpha_{\;\beta\gamma}$ the connection
$\bar{\Gamma}^\alpha_{\;\beta\gamma}$ is nonsymmetric, and
$$
\bar{\Gamma}^\alpha_{\;\beta\gamma} -
\bar{\Gamma}^\alpha_{\;\gamma\beta} =
T^\alpha_{\;\beta\gamma}                        \eqno(1)
$$
If one introduce the metricity condition
$\bar{\nabla}_\mu g_{\alpha\beta} = 0$ where the covariant derivative
$\bar{\nabla}_\mu$ is constructed on the base of
$\bar{\Gamma}^\alpha_{\;\beta\gamma}$ then the following solution for
connection
$\bar{\Gamma}^\alpha_{\;\beta\gamma}$ can be easily found
$$
\bar{\Gamma}^\alpha_{\;\beta\gamma} = {\Gamma}^\alpha_{\;\beta\gamma} +
K^\alpha_{\;\beta\gamma}                         \eqno(2)
$$
where ${\Gamma}^\alpha_{\;\beta\gamma}$ is standard symmetric Christoffel
symbol and $K^\alpha_{\;\beta\gamma}$ is contorsian tensor
$$
K^\alpha_{\;\beta\gamma} = \frac{1}{2} \left( T^\alpha_{\;\;\beta\gamma} -
T^{\;\alpha}_{\beta\;\gamma} - T^{\;\alpha}_{\gamma\;\beta} \right)
                                                       \eqno(3)
$$
It is convinient to divide the torsion field into three irreducible components
that are: the trace $T_{\beta} = T^\alpha_{\;\beta\alpha}$, the pseudotrace
$S^{\nu} = \varepsilon^{\alpha\beta\mu\nu}T_{\alpha\beta\mu}$ and the tensor
$q^\alpha_{\;\beta\gamma}$, which satisfy the conditions
$$
q^\alpha_{\;\beta\alpha} = 0,\;\;\;\;\; \;\;\;\;
\varepsilon^{\alpha\beta\mu\nu}q_{\alpha\beta\mu} =0
$$

Then the torsion field can be written in the form
$$
T_{\alpha\beta\mu} = \frac{1}{3} \left( T_{\beta}g_{\alpha\mu} -
T_{\mu}g_{\alpha\beta} \right) - \frac{1}{6} \varepsilon_{\alpha\beta\mu\nu}
S^{\nu} + q_{\alpha\beta\mu}                \eqno(4)
$$

Now we consider the Dirac field $\psi$
in an external gravitational field with torsion.
A various aspects of the interaction of the Dirac field with torsion
have been discussed in the literature (see, for example, [1,2, 15 - 18]).
It is well known that the standard way to introduce the minimal interaction
with external fields require the substitution of the partial
derivatives $\partial_\mu$ by the covariant ones. The covariant derivatives
of the spinor field $\psi$ are defined as follows
$$
\bar{\nabla}_\mu \psi = \partial_{\mu}\psi + \frac{i}{2}w_\mu^{\; a b}
\sigma_{a b}\psi
$$
$$
\bar{\nabla}_\mu \bar{\psi} = \partial_{\mu}\bar{\psi} -
\frac{i}{2}w_\mu^{\;a b}\bar{\psi}\sigma_{a b}       \eqno(5)
$$
where $w_\mu^{\;a b}$ are the components of spinor connection. We use the
standard representation for the Dirac matrices (see, for example, [20]).
$$
\beta = \gamma^0 = \left(\matrix{1 &0\cr 0 &-1\cr} \right)
$$
$$
\vec{\alpha} = \gamma^0 \vec{\gamma} =
\left(\matrix{0 &\vec{\sigma}\cr 0 &\vec{\sigma}\cr} \right)
$$
$$
\gamma_5 = \gamma^0 \gamma^1 \gamma^2 \gamma^3, \;\;\;\;
\sigma_{a \; b} = \frac{i}{2}(\gamma_a \gamma_b - \gamma_b \gamma_a)
$$
The verbein $e_\mu^a$ obey the equations $e_\mu^a e_{\nu a} = g_{\mu\nu}$,
$e_\mu^ae^{\mu b} = \eta^{ab}$ and $\eta^{ab}$ is the Minkowsky metric. The
gamma matrices in curved space - time are introduced as $\gamma^\mu =
e_a^\mu \gamma^a$ and obviously satisfy the metricity condition
$\bar{\nabla}_\mu \gamma^{\beta} = 0$ .
The condition of metricity enables us to find the explicit expression
for spinor connection which agree with (2).
$$
w_\mu^{\;a b} = \frac{1}{4} (e_\nu^b \partial_\mu e^{\nu a} -
e_\nu^a \partial_\mu e^{\nu b}) + \bar{\Gamma}^\alpha_{\;\nu\beta}
(e^{\nu a}e_\alpha^b - e^{\nu b}e_\alpha^a)         \eqno(6)
$$

If the metric is flat, then from (6) follows
$$
w_\mu^{\; a b} = K^\alpha_{\;\nu\beta}
(e^{\nu a}e_\alpha^b - e^{\nu b}e_\alpha^a)                           \eqno(7)
$$

The action of spinor field minimally coupled with torsion have the form
$$
S = \int d^4 x \;e \; \{ \frac{i}{2}\bar{\psi}\gamma^\mu \bar{\nabla}
_\mu \psi - \frac{i}{2}\bar{\nabla}_\mu\bar{\psi}\gamma^\mu\psi +
m\bar{\psi}\psi \}                                               \eqno(8)
$$
where $m$ is the mass of the Dirac field and
$e = \det \parallel e_\mu^a\parallel$. Further we shall consider only the
torsion effects and therefore restrict ourselves by the only special case
of flat metric. So we put $g_{\mu\nu} = \eta_{\mu\nu}$ but keep
$T^\alpha_{\;\beta\gamma}$ arbitrary. The expression (8) can be rewritten in
the form
$$
S = \int d^4 x\{i\bar{\psi}\gamma^\mu(\partial_\mu+\frac{i}{8}\gamma_5
S_\mu)\psi+m\bar{\psi}\psi\}                                     \eqno(9)
$$
One can see that the spinor field minimally interact only with the
pseudovector $S_\mu$ part of the torsion tensor. The nonminimal interaction is
more complicated.
There are strong reasons to introduce the nonminimal coupling of the form
$$
S = \int d^4 x\{i\bar{\psi}\gamma^\mu(\partial_\mu
+i\eta_1\gamma_5S_\mu+i\eta_2T_\mu)\psi+m\bar{\psi}\psi\}        \eqno(10)
$$
Here $\eta_1,\eta_2$ are the dimensionless parameters of the
nonminimal coupling of spinor fields with torsion. The minimal
interaction corresponds to the values $\eta_1 = \frac{1}{8},\;\;\eta_2 = 0$.

The introduction of the nonminimal interaction looks artificial.
Within the classical theory one can explain the use of a nonminimal
action only as an attempt to explore the more general case. However
the situation is different in quantum region where the nonminimal
interaction is the necessary condition of consistency of the theory.
The reason is following. It is well-known that the interaction of
quantum fields leads to the divergences and therefore the
renormalization is needed. As it was shown in
[12, 13], the requirement of the multiplicative
renormalizability makes us to introduce the nonminimal interaction of torsion
with spinor and scalar fields.

\section{The equation of motion for spinor field in the
weakrelativistic approximation}

Let us consider the spin $\frac{1}{2}$ particle in an external torsion
and electromagnetic fields. The equation of motion follows from (10)
with the usual  electromagnetic addition.
$$
i \hbar\frac{\partial \psi}{\partial t} = \{ c\vec{\alpha}\vec{p}-
e\vec{\alpha}\vec{A} - \eta_1 \vec{\alpha}\vec{S}\gamma_5 -
\eta_2 \vec{\alpha}\vec{T}+
$$
$$
+ e\Phi + \eta_1 \gamma_5 S_0 + \eta_2 T_0 + m c^2 \beta \}\psi
                                                              \eqno(11)
$$
Here the dimensional constants $\hbar$ and $c$ are taken into account,
$A_\mu = (\Phi, \vec{A}),\;\; T_\mu = (T_0, \vec{T}),\;\; S_\mu = (S_0,
\vec{S})$

Following the standard prosedure we write (see, for example, [20])
$$
\psi = \left(\matrix{\varphi\cr\chi\cr}\right)
exp( \frac{imc^2t}{\hbar})                                  \eqno(12)
$$

Within the weakrelativistic approximation $\chi \ll \varphi$. From
equations (11), (12) it follows that
$$
(i\hbar\frac{\partial}{\partial t}-
\eta_1 \vec{\sigma}\vec{S} - e\Phi - \eta_2 T_0)\varphi=
$$
$$
= (c\vec{\sigma}\vec{p} - e\vec{\sigma}\vec{A} - \eta_1 S_0 -
\eta_2 \vec{\sigma}\vec{T})\chi
                                                          \eqno(13a)
$$
and
$$
(i\hbar\frac{\partial}{\partial t}-
\eta_1 \vec{\sigma}\vec{S} - e\Phi - \eta_2 T_0 + 2mc^2 ) \chi =
$$
$$
= (c\vec{\sigma}\vec{p} - e\vec{\sigma}\vec{A} - \eta_1 S_0 -
\eta_2 \vec{\sigma}\vec{T})\varphi
                                                          \eqno(13b)
$$
Now we keep only the term $2mc^2\chi$ in the left side of (13b) and
then it is possible to find $\chi$ from (13b). In the leading order in
$\frac{1}{c}$ we meet the following eqation for $\varphi$.
$$
i \hbar\frac{\partial \varphi}{\partial t} =
\{ \eta_1 \vec{\sigma}\vec{S} + e\Phi + \eta_2 T_0 +
$$
$$
+ \frac{1}{2mc^2\chi} (c\vec{\sigma}\vec{p} - e\vec{\sigma}\vec{A} -
\eta_1 S_0 - \eta_2 \vec{\sigma}\vec{T} )^2 \} \varphi      \eqno(14)
$$
The last equation is easily rewritten in the Scrodinger form
$$
i\hbar\frac{\partial \varphi}{\partial t} =
\hat{H} \varphi                                                   \eqno(15)
$$
where the Hamiltonian have the form
$$
\hat{H} = \frac{1}{2m} \vec{\pi}^2 + B_0 + \vec{\sigma}\vec{Q}
$$
$$
\vec{\pi} = \vec{P} - \frac{e}{c}\vec{A} - \frac{\eta_2}{c}\vec{T} -
\frac{\eta_1}{c}\vec{\sigma}S_0
$$
$$
B_0 = \frac{e}{\Phi} + \eta_2 T_0 - \frac{1}{mc^2}\eta_1^2 S_0^2
$$
$$
\vec{Q} = \eta_1\vec{S} + \frac{\hbar}{2mc}(e\vec{H} + \eta_2\;rot\vec{T})
                                                                  \eqno(16)
$$

Here $\vec{H} = rot\vec{A}$ is the magnetic field strength.
The equation of (15), (16) is the analog of the Pauly equation in the
case of external torsion and electromagnetic field.

The expression for the Hamiltonian (16) indicate on the physical
effects of the torsion field, that is especially clear if compared with
the electromagnetic terms. For example, the quantity
$T_0$ looks like scalar potential $\Phi$,  $\vec{T}$ looks like vector
potential $\vec{A}$. The quantities $\vec{S}$ and $rot\;\vec{T}\;$ may
play the role of the magnetic field. However there is some difference
between torsion and electromagnetic sectors. The term
$- \frac{1}{mc}\eta_1 S_0 \vec{p}\vec{\sigma}\;$ does not have the
analogies in quantum electrodynamics.

\section{The equation of motion for the particle with spin $\frac{1}{2}$ in an
external torsion field.}

If we consider (16) as the Hamiltonian operator of some quantum
particle, then the corresponding classical energy have the form
$$
H = \frac{1}{2m} \vec{\pi}^2 + B_0 + \vec{\sigma}\vec{Q}
                                                              \eqno(17)
$$
where $\vec{\pi}, B_0, \vec{Q}$ are defined by (16) and
$\vec{\pi} = m \vec{v}$. Here $\vec{v} = \dot{\vec{x}}$
is the velocity of the particle. From (17) it follows the the expression for
the
 canonical conjugated momenta $\vec{p}$.
$$
\vec{p} = m\vec{v} + \frac{e}{c}\vec{A} + \frac{\eta_2}{c}\vec{T} +
\frac{\eta_1}{c}\vec{\sigma}S_0                               \eqno(18)
$$
One can consider $\vec{\sigma}$ as the coordinate of internal degrees of
freedom, corresponding to spin.

Let us now perform the canonical quantization of the theory. To make this we
introduce the operators of coordinate $\hat{x}_i$, momenta $\hat{p}_i$ and
spin $\hat{\sigma}_i$ and input the equal - time commutation relations of
the following form:
$$
\left[\hat{x}_i, \hat{p}_j\right] = i\hbar \delta_{ij}, \;\;\;\;\;
\left[\hat{x}_i, \hat{\sigma}_j \right] =
\left[\hat{p}_i, \hat{\sigma}_j\right] = 0,
$$
$$
\left[\hat{\sigma}_i,\hat{\sigma}_j \right] = 2i\varepsilon_{ijk}
\hat{\sigma}_k
                                                              \eqno(19)
$$
The Hamiltonian operator $\bar{H}$ which corresponds to the energy (17) is
easily
constructed in terms of the operators $\hat{x}_i, \hat{p}_i, \hat{\sigma}_i$
and then these operators yield the equations of motion
$$
i\hbar \frac{d\hat{x}_i}{dt} = \left[\hat{x}_i, H \right],
$$
$$
i\hbar \frac{d\hat{p}_i}{dt} = \left[\hat{p}_i, H \right],
$$
$$
i\hbar \frac{d\hat{\sigma}_i}{dt} = \left[\hat{\sigma}_i, H \right],
                                                                \eqno(20)
$$

After the computation of the commutators in (20) we obtain the
explicit form of the operators equations of motion. Now we can omit
all the terms which vanish when $\hbar \rightarrow \; 0$. Then the
classical equations arise which can be interpreted as the
(quasi)classical  equations of motion for the particle in
external torsion and electromagnetic fields. Note that the operator
arrangement problem is irrelevant because of the use of
$\hbar \rightarrow \; 0$ limit. The straightforward calculations
lead to the equations
$$
\frac{d\vec{x}}{dt} = \frac{1}{m} \left( \vec{p} - \frac{e}{c}\vec{A} -
\frac{\eta_2}{c}\vec{T} - \frac{\eta_1}{c}\vec{\sigma}S_0 \right) = \vec{v},
                                                             \eqno(21a)
$$
$$
\frac{d\vec{v}}{dt} = e\vec{E} + \frac{e}{c}\left[ \vec{v}\times\vec{H} \right]
+ \frac{\eta_2}{c}\left[ \vec{v}\times rot\;\vec{T} \right] -\eta_2\; grad\;T_0
-
$$
$$
- \frac{\eta_2}{c}\frac{\partial\vec{T}}{\partial t} -
\eta_1\left(\vec{\sigma}\cdot\nabla \right)\vec{S} -
\eta_1\left[ \vec{\sigma}\times rot\vec{S} \right]
- \frac{\eta_1}{c}\vec{\sigma}\frac{\partial S_0}{\partial t} +
$$
$$
+ \frac{\eta_1}{c} \{ \left(\vec{v}\cdot\sigma\right) grad S_0 -
\left(\vec{v}\cdot grad S_0 \right)\vec{\sigma} \}
+\frac{1}{mc^2}\;\eta_1^2\; grad (S_0^2) -
\frac{\eta_1}{c}S_0\frac{d\vec{\sigma}}{dt},                   \eqno(21b)
$$
$$
\frac{d\vec{\sigma}}{dt} = \left[ \vec{R}\times\vec{\sigma} \right]
$$
$$
\vec{R} = \frac{2\eta_1}{\hbar}\left[ \vec{S} - \frac{1}{c}\vec{v}S_0 \right]
+ \frac{e}{mc}\vec{H} + \frac{\eta_2}{mc}\; rot\vec{T}
                                                             \eqno(21c)
$$

Here $\vec{E}$ is the strength of the external electric field. Equations (21)
contain the torsion - dependent terms which have the same symmetries as
the usual electromagnetic terms. Really, the $T_\mu$ dependent terms are in a
perfect analogy with the $A_\mu$ dependent terms. However the equations (21)
contains some terms which have a qualitatively new structures.
All this terms contains
$S\mu$, that is more relevant part (with respect to the interaction with the
matter fields) of the torsion tensor.
Thus we see that standard claim conserning magnetic field analogy of torsion
effects is not completely correct, and there exist serious difference between
magnetic field and torsion.

\section{Possible experimental investigations of the torsion field}

Let us now consider the Schrodinger equation (15) with the Hamiltonian
operator (16) with the vanishing electromagnetic field. How can the torsion
field
manifest itself? It is evident that the effect of torsion field can modify
the particles spectrum. This modifications have the similar form to the ones
which arise in the elecromagnetic field. At the same time another modifications
are possible due to the qualitatively new terms like
$\;\frac{1}{mc}\eta_1 S_0 \vec{p}\vec{\sigma}\;$ in (16).

It is natural to suppose that the possible interaction with torsion is feeble
enough and therefore one can consider it as some perturbation. This
perturbation
may leads to the splitting of the known spectral lines and hence one can hope
to find the torsion display within the spectral analysis experimants.
In particular one can expect the splitting of the spectral lines even for the
more simple hydrogen atom. Now we consider the particular case of
$\;T_\mu = 0,\;\;
S_\mu = const\;$ and estimate the possible spectrum modifications. In this
particular case Hamiltonian operator is
$$
\hat{H} = \frac{1}{2m}\hat{\pi}^2 + \eta_1 \left( \hat{\vec{S}} -
\frac{1}{2mc}\hat{S}_0\hat{\vec{p}} - \frac{1}{2mc}\hat{\vec{p}}
\hat{S}_0 \right)\cdot \hat{\vec{\sigma}}
$$
$$
\vec{\pi} = \vec{p} - \frac{e}{c}\vec{A}                \eqno(22)
$$

In the framework of the nonrelativistic approximation $|\vec{p}| \ll
mc$ and hence the second $S_0$ dependent term in the bracets can be omitted.
The
remaining term $\eta_1\vec{S}\vec{\sigma}$ allows the standard
interpretation and gives the contribution $\pm \eta_1 S_3$ into the
spectrum. Thus, if the $S_3$ component of the torsion tensor is not
equal to zero, the energy level is splitted into two sublevels with
the difference $2 \eta_1 S_3$. If now the week transversal magnetic
field is switched on then the cross between the new levels will arise
and energy absorbtion takes place at the magnetic field frequency $w =
\frac{2\eta_1}{S_3}$. Note that the situation is typical within the
magnetic resonance experiments, however in the case this effect arise
due to the torsion, but not magnetic field effects.
It is natural this effect as the torsion resonance. Taking into
account the previous consideration we arise at the conclusion that the
described effect can be explored at different scales: torsion -
induced spin resonance in atoms, the torsion electron resonance and
the torsion nuclear resonance in a medium. Note that the experiments
related with torsion induced splitting of the energy levels was recently
considered in [23].

The next kind of the possible experinments related with torsion are
the ones which deal with the particle equation of motion (21). Let the
electromagnetic field is absent.

Then, according to (21) the interaction with torsion twists the
particle trajectory and therefore any charged particles may be the
sourse of the electromagnetic radiation. Then the structure of the radiation
field enables one to to look for the torsion effects.

Of course, some evident effects like the precession of spin in an
external torsion field also follows from (21). This effect have been
already described in Ref.'s [1,2,15,16]. It is interesting that from
(21c) follows that the direction of precession of spin depends on the
velocity of the particle. Therefore even the week torsion field may
violate the observable precession of spin in a magnetic field
at high temperature.

To obtain the complete picture of the torsion influence to the energy
spectrum modifications as well as to the radiation of charged
particles it is necessary to make the detailed and systematic
investigation of the equations (15), (21) solutions in the case of
a various external field structures. The main difficulty is that there
are no any experimental data for the values of coupling
constants. That is why it is impossible to give the numerical estimate
for the mentioned physical effects.
Note that the inflationary cosmological model with torsion predict a tiny
value of torsion field, which have to be very slowly variing in a
modern epoch [21]. Thus there are some reasons to look for the evidence of some
weak
global (effectively) axial vector in the Universe, and try to give the upper
bound for torsion field from any modern experiments.

 From the results of Ref's [12,13,7] it follows that the interaction of
torsion with matter fields is essentially weakened at low energies due to
quantum effects. That is why we have a very small hope to observe torsion in a
low - energy experiments. On the other hand even the very weak interaction with
torsion may be responsible for some symmetry violation because of the
pseudovector nature of the vector $S_\mu$.
Indeed such an effects are essentially related with high-energy physics and
our consideration have to be extended. In any case the results of the above
analysis may be useful in qualitative understanding of the structure of
torsion - matter interaction.

\section{Acknowledgments}
Authors are appreciate  Professors G.Cognola, T.Kinoshita, I.B.Khriplovich,
T.Muta  and
S.Zerbiny for useful conversations.
One of the authors (I.Sh.) wish to thanks Particle Physics Group
at Hiroshima University and Theory Division at KEK for kind hospitality.

\end{document}